\documentclass[12pt,oneside]{amsart}

      \usepackage{graphicx}
      \usepackage{amssymb}

      \theoremstyle{plain}

      \theoremstyle{definition}

      \theoremstyle{remark}

      \makeatletter
      \def\@setcopyright{}
      \def\serieslogo@{}
      \makeatother
   
\begin{document}

%



   \author{Brandon K. Wiggins$^{1,2}$}
   \address{Brandon K. Wiggins}
   \email{brandonwiggins@byu.edu}


   \author{Joseph M. Smidt$^1$}


   \author{Daniel J. Whalen$^3$}


   \author{Wesley P. Even$^1$}

   \author{Victor Migenes$^2$}

   \author{Chris L. Fryer$^1$}


   \title[Hypernovae and Pair-Instability Supernovae]{Assessing the Observability of \\ Hypernovae and Pair-Instability Supernovae in the Early Universe}


   \begin{abstract}

The era of the universe's first (Population III) stars is essentially unconstrained by observation. Ultra-luminous and massive stars from this time altered the chemistry of the cosmos, provided the radiative scaffolding to support the formation of the first protogalaxies, and facilitated the creation and growth of now-supermassive black holes. Unfortunately, because these stars lie literally at the edge of the observable universe, they will remain beyond the reach of even the next generation of telescopes such as the James Webb Space Telescope and the Thirty-Meter Telescope. In this paper, we provide a detailed primer to supernovae modeling and the first stars to make our discussion accessible to those new to or outside our field. We review recent work of the Los Alamos Supernova Light Curve Project and Brigham Young University to explore the possibility of probing this era through observations of the spectacular deaths of the first stars.  We find that many such brilliant supernova explosions will be observable as far back as $\sim 99$ \% of the universe's current age, tracing primordial star formation rates and the locations of their protogalaxies on the sky. The observation of Population III supernovae will be among the most spectacular discoveries in observational astronomy in the coming decade.

   \end{abstract}



      \thanks{[1] Los Alamos National Laboratory, Los Alamos, NM}
   \thanks{[2] Brigham Young University, Provo, UT}

   \thanks{[3]  Max Planck Institute for Theoretical Astrophysics}
   \thanks{Los Alamos Unlimited Release number: LA-UR-14-24307}

   \dedicatory{To appear in the Journal of the Utah Academy}

   \date{\today}


   \maketitle



   \section{Introduction}

Because the universe had a beginning, there must have been a first star. Supercomputer simulations now show that the first stars probably formed at redshift $z \sim 20$, or only $200 - €"400$ million years after the Big Bang in an event now called ``Cosmic Dawn.'' But star formation during this time was very different from what we observe in the universe today. The first stars formed in small pregalactic structures known as cosmological halos, in pristine hydrogen and helium gases that were devoid of the heavier elements and dust that are ubiquitous in star-forming clouds today. A hypothetical telescope image of this time would not contain spiral and elliptical galaxies, but individual stars sprinkled through the universe's invisible dark-matter filamentary structure and slowly collecting in small stellar communities.

Because these stars were discovered most recently, they are called Population III (Pop III) stars; Pop III stars were created from only hydrogen and helium, and they would have had some unfamiliar properties. Clouds with this simple composition do not cool well, and this gave rise to truly gigantic stars that were hundreds of times more massive than the Sun and tens of millions of times more luminous (e.g., Hirano et al. 2014). A few may have been a hundred thousand times the mass of the Sun and may have been the precursors of supermassive black holes (Wise et al. 2008; Regan and Hehnelt 2009a). In a sense, one can think of this as the universe's very own \textit{Jurassic era} where it manufactured \textit{monsters.} These cosmological ``dinosaurs'' became extinct long ago as the first stars spewed heavier elements throughout the cosmos in spectacular supernova explosions. Chemically enriched interstellar clouds cool more rapidly and so are unable to grow to such large masses before collapsing to form stars. Thus, the modern cosmos is incapable of making the gigantic stars that may have been found in the primeval universe. Remains of this period likely persist today in the chemical composition of ancient, dim stars of the halos of galaxies.

It isn'€™t merely the curious character of the first stars that fascinates cosmologists. The first stars populated the cosmos with heavy elements, allowing for the later formation of planets and life. Their light also gradually transformed the universe from a cold, dark, featureless void into the vast, hot, transparent cosmic web of galaxies we observe today (Bromm et al. 2009). These stars and this period are key in resolving standing cosmological mysteries such as how super-massive black holes billions of times more massive than the Sun appeared less than a billion years after the Big Bang (Moretti et al. 2014). Primordial stars also populated the first primitive galaxies, which will be principal targets of the James Webb Space Telescope (JWST), which is slated to launch in 2018 (Bromm et al. 2009). Understanding the nature of the first stars is crucial to predicting the luminosities and spectra of primeval galaxies.

Unfortunately, these stars are beyond the reach of current observations. Individual primordial stars will not even be visible to the next generation of 30-meter class telescopes or space missions because they literally lie near the edge of the observable universe, when it was only 200 million years old. But the first stars may have died in luminous supernova explosions, and these spectacular events may be visible to upcoming instruments. Indeed, supernovae in the local universe can outshine their entire host galaxies and have now been detected as far away as 10 billion light years. With the coming telescopes, it may be possible to see even more distant explosions, which took place at Cosmic Dawn. These ancient supernovae (SNe) may offer the first observational constraints on this important but elusive and distant epoch of cosmic history.

To interpret the findings of future observations, cosmologists rely on predictions regarding the character, the luminosity, and the frequency of ancient SNe. In particular, cosmologists must know whether Pop III SNe are sufficiently luminous to be observable to upcoming instruments. Computer simulations are employed to provide models for the light curves (total luminosity as a function of time) and the spectra of these spectacular events. These will be used by observers to identify and characterize ancient SNe. Some types of primordial SNe have been modeled in the literature (see, e.g., Whalen et al. 2013a, 2013b, 2014b). In this study, we will investigate the observability of Pop III pair-instability supernovae (PI SNe) and hypernovae (HNe), which are types of brilliant supernovae explosions of very large stars. This era of the first stars in the universe may well have been the epoch of its most luminous supernovae if the first stars were sufficiently massive.

\begin{center}

\textsc{1.1 how big was big?}
\end{center}

Like deaths of stars in the local universe, the character (including the luminosity) of the explosive deaths of the first stars was primarily dependent upon their masses. The literature, however, has not always been in agreement on how massive Pop III stars actually were. Some studies suggest that they were tens of solar masses (e.g., Hosokawa et al. 2011), but others have found that some may have had masses of 500–600 times that of the Sun (e.g., Hirano et al. 2014). These estimates are largely based on simulations that attempt to follow the collapse and accretion of gas onto a protostar; however, no high-resolution simulation has evolved the protostar for more than a 1000 years, while the time between the formation and evaporation of its accretion disk may take millions of years (see Whalen 2012). Consequently, simulations cannot yet constrain the masses of the first stars.

Some properties of primordial stars can be inferred from the chemical abundances found in old halo stars in our own galaxy (see, e.g., Christlieb et al. 2002, Frebel et al. 2005). When the first stars died, ashes from their explosions may have been taken up in the formation of the next generation of stars. Joggerst et al. (2009) found that the cumulative nucleosynthetic yields of 15- to 40-solar mass Pop III supernovae are a good match to the elemental abundances measured in the extremely metal-poor stars to date. This finding would apparently contradict the results of simulations predicting Pop III stars of 100s of solar masses; however, ``stellar archeology'' as this study is called, is still in its infancy because of small sample sizes and the fact that the very metal-poor stars found so far reside in the galactic halo instead of the galactic nucleus, where most second-generation stars would be expected (Hirano et al. 2014). Most sources place Pop III stars as having between 50 to 500 solar masses. In modeling PI SNe and HNe, we are exploring the widely accepted paradigm that the first stars were very large.

\begin{center}

\textsc{1.2 hearts of darkness and hearts of antimatter}
\end{center}

While less massive Pop III stars ended their lives as white dwarfs, neutron stars, or black holes, in some cases with core-collapse supernova explosions, massive Pop III stars would have died more exotically and spectacularly. Rakavy and Shaviv (1967) proposed that stars larger than 140 solar masses die in PI SNe. In these scenarios, core temperatures of the star during oxygen burning exceed about a billion degrees Kelvin, and thermal photons are converted into electron-€"positron pairs. This robs the core of radiation pressure, causing the core to contract and its temperature to rise. Explosive oxygen and silicon burning result. Whereas less massive stars die with the collapse of their core, these stars' cores go up in a powerful nuclear explosion some 100 times more energetic than core-collapse supernovae. The energy release completely unbinds the star in a brilliant explosion. PI SNe synthesizes up to 40 solar masses of $^{56}$Ni whose subsequent radioactive decay can power the luminosities of these SNe for up to 3 years (Whalen et al. 2013a). The idea of a star with a core so hot that it creates antimatter may seem like science fiction, but a few PI SN candidates have now been found in the local universe (see Pan et al. 2012b).

Previous studies have examined the visibility of 140- to 260-solar mass Pop III PI SNe to future telescopes (e.g., Whalen et al. 2013a), but new work has shown that rapidly rotating stars can encounter the pair instability at somewhat lower masses (Chatzopoulos and Wheeler, 2012). Rapid rotation mixes the star'€™s layers and effects homogenous nuclear fusion throughout the star. This leads to a buildup of a larger oxygen core, which can trigger the pair instability at lower stellar masses. Lower-mass Pop III stars would have been much more common than their very high mass counterparts, and so we will investigate the observability of these newly discovered SNe. 

Primordial stars between about 25 and 60 solar masses may die as HNe. These supernovae were not as energetic as PI SNe but were still luminous and likely sufficiently energetic to be observed in upcoming surveys. Although HNe have been observed (e.g., Nomoto et al. 1998), their central engines are not yet well understood. A prominent model is the collapsar model (Woosley 1993), in which the core of a rapidly rotating star collapses to a black hole surrounded by an accretion disk. Rapid infall onto the black hole drives a relativistic jet into the outer layers of the star, which are still collapsing. The jet breaks through these outer layers in a highly asymmetric explosion that is very luminous along the line of sight of the jet. Fallback onto the star's newly formed and rotating ``€œheart of darkness''€ powers these spectacular, beamed, events.

Finding these early cosmic explosions will open our first direct window on the primeval universe. If Pop III SNe are detected in sufficient numbers, the fact that distinct types of supernovae occur over different intervals of stellar mass could soon allow observers to determine how Pop III stars were distributed in mass. To some degree, the mass of the progenitor can be inferred from the light curve of its explosion. Primordial SN rates could also constrain star formation rates in the early universe. These events could also pinpoint the positions of primitive galaxies on the sky, especially if the galaxy is too dim to otherwise be detected. Likewise, the failure to detect Pop III SNe could imply lower star formation rates or less massive Pop III stars, either of which would also be important discoveries. The observation of Pop III supernovae by future telescopes will be a landmark achievement in astronomy in the coming years.

   \section{Method}
  
A thorough discussion of the Los Alamos Supernova Light Curve Project can be found in Frey et al. (2013). This source may be consulted for more technical details regarding our study. 

We have modeled 12 PI SNe of rotating stars with masses from 90 to 140 solar masses in increments of 5 solar masses. We have also simulated six HNe. Observationally, hypernovae have been inferred to have explosion energies from 10 to 50 foe (Smidt et al. 2014), where 1 foe $= 10^{51}$ erg is the typical energy of a Type II SN. To explore the parameter space of likely Pop III HNe, we consider 10-, 22-, and 52-foe explosions of two hypernovae progenitors of 25 and 50 solar masses, respectively.

\begin{center}

\textsc{2.1 following collapse and explosion}
\end{center}

Our procedure differs in the modeling of PI SNe and HNe. We will describe the modeling of HNe and PI SNe in turn. Figure 1 provides a schematic which may assist the reader as we discuss our method below.

\begin{figure}
\includegraphics[width = \columnwidth]{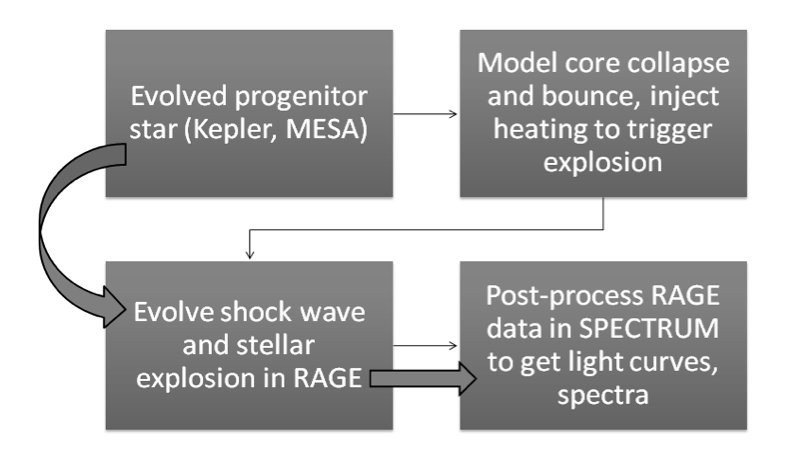}
\caption{\footnotesize Schematic of our simulation process between various codes. To model hypernovae, we used the track indicated with the thin arrows as we needed to model core collapse and bounce and inject energy to trigger the explosion. To model pair-instability supernovae (PI SNe), we followed the track indicated with the thicker arrows as the explosion in PI SNe is emergent from the stellar evolution code. All simulations required RAGE and post-processing with SPECTRUM.}
\end{figure}

The energy and luminosity of a SN can depend strongly on the structure of the star prior to the explosion. To obtain the final profiles for the progenitor star, we evolve it from birth to the onset of collapse in the Kepler (Weaver et al. 1978) or MESA (Modules for Experiments in Stellar Astrophysics) (Paxton et al. 2013) stellar evolution codes. Because hypernovae are observed to be Type Ib/c SNe with no hydrogen lines in their spectra, we then strip off the hydrogen layer from both stars (recall we only have a 25- and a 50-solar mass progenitor) before transferring it to a one-dimensional Lagrangian core collapse code (Fryer 1999). This code follows the collapse of the star through the time when the core stops contracting and bounces, at which point energy due to neutrino absorption is artificially injected into the inner layers (the innermost 15 cells) to drive a range of explosion energies (see Young and Fryer 2007 for additional details on this code). We note that both stellar evolution and explosive nuclear burning must be modeled with extensive nuclear reaction networks that are self-consistently coupled to hydrodynamics to capture both energy production and nucleosynthetic yields. After nuclear burning is complete, which takes a few hundred seconds in the frame of the star, the HN is then evolved in Los Alamos' radiation hydrodynamics code RAGE (Radiation Adaptive Grid Eularian) to follow the evolution of the shock wave as it travels through the star, bursts through its surface, and expands into the surrounding medium. The Kepler calculations require on the order of 24 hours on Los Alamos National Laboratory (LANL) platforms. 

Our pair-instability progenitors were similarly modeled in Kepler and MESA, but there is no need to model core collapse and bounce or inject energy to drive the explosion. Pair production, core contraction, and explosive oxygen and silicon burning in PI SNe are emergent features of the stellar evolution model and do not have to be artificially triggered. Nuclear burning is usually finished in $10–30$ seconds in PI SNe, after which it is transferred to RAGE. 

Although RAGE can follow the evolution of SN flows and radiation coming from them, it cannot calculate light curves or spectra for the explosion (which are what would actually be observed by astronomers). To calculate the observational signatures of these explosions, we post process snapshots of the flow from RAGE with the Los Alamos SPECTRUM code. SPECTRUM calculates luminosities for the SN in 13,900 wavelength bins, which can then be summed to create light curves. SPECTRUM uses the LANL OPLIB (OPacity LIBrary) opacity database (Magee et al. 1995) to determine from which regions of the flow photons can escape to an external observer. SPECTRUM can also calculate the intensities of emission and absorption lines and take into account redshifting and blueshifting of photons due to relativistic expansion of SN ejecta. Spectra in the frame of the SN at very early times must then be cosmologically redshifted and subtracted by absorption in the intervening gas to determine light curves for the event in the Earth frame (``cosmological redshifting'' refers to the stretching of photon wavelengths by the expansion of the universe over cosmic time as light from the event reaches Earth). The SPECTRUM runs required to calculate a single light curve require as many CPU hours as a RAGE run but can be executed in much shorter wall clock times because they can be run in parallel (usually only 1–2 days are required per light curve).

All the simulations in this paper are performed in one dimension and therefore exclude multidimensional effects that can break spherical symmetry such as hydrodynamical instabilities, magnetic fields, and turbulence. Our simulations also cannot capture orientational effects, which are thought to be important for some SNe. In some cases, mixing and dredging increases the luminosities of some events. Over large enough sample sizes, however, the simulations do give results sufficiently robust to estimate detection limits for Pop III SNe in redshift. Our SPECTRUM calculations (and the LANL OPLIB opacities on which they rely) also assume that matter is in local thermodynamic equilibrium, which may break down at later times when the supernova ejecta becomes diffuse. Our one-dimensional models of HNe also treat these highly asymmetric explosions as spherical events, but we inject enough explosion energy over the entire sphere to approximate the energy emitted along just the jet. Our simulations therefore should produce reasonable estimates of HN luminosities.

Our procedure for modeling hypernovae may raise the question as to how sensitive are our results to be to the magnitude of the explosion energy injected through neutrino absorption. Although no alternative to injecting energy over the whole sphere is possible in a one-dimensional hypernova simulation, we must consider how sensitive our results are to small variations in explosion energy. Our experiment is naturally set up to bracket this, as our six hypernovae explosions are created by varying explosion energy on only two progenitors from 10 to 52 foe. The variation in our results over a given progenitor will directly shed light on this issue. In general, we find that a twofold increase in explosion energy results in a $\sim$twofold increase in peak luminosities. 

\begin{center}

\textsc{2.2 evolution of the supernova remnant in rage}
\end{center}

	After explosive nuclear burning is complete, output is fed to RAGE to model the radiation hydrodynamical evolution of the shock wave. The code captures the effects of radiating matter, which is in turn heated and accelerated by light. As the shock bursts through the surface of the star in an event called ``shock breakout'', light trapped behind the shockwave streams freely into space. The accelerating shockwave heats material surrounding the star to white-hot temperatures, setting it ablaze with light (see Figure 2). The grand effect is a sharp, brilliant pulse of light. The remnant can rebrighten at later times as radioactively decaying $^{56}$Ni is exposed in the expanding remnant and heats the stellar material. All of these effects are modeled in RAGE.

\begin{figure}
\includegraphics[width = 280 pt]{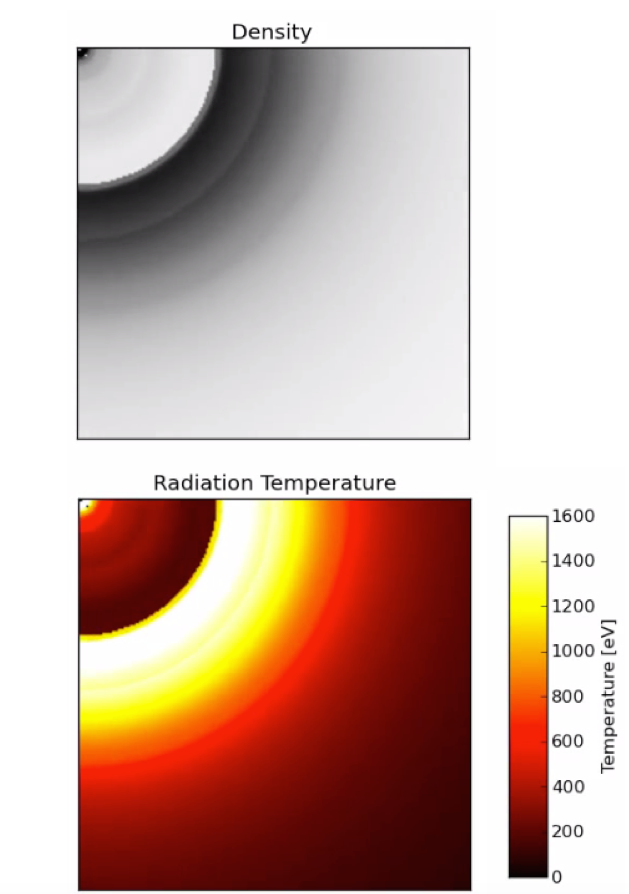}
\caption{\footnotesize Snapshot of a RAGE hypernova simulation sometime following the moment when the shockwave breaks through the surface of the star and plows through the ambient medium. RAGE simulations are carried out in one dimension, so the extra dimension visualized here serves only to emphasize structure and provide a feel for the phenomenon. In the upper panel, relative density is indicated, black being regions with the most material. At this stage, the star has been almost completely disrupted into a ring (or a sphere in three dimensions) of glowing material. Matter has fallen back onto the central black hole (upper lefthand corner), which is radiating as observed in the bottom frame. Note that though the scale bar in the lower panel only indicates temperature up to 1600 eV, regions of this panel are tens of thousands of eV.}
\end{figure}

The Los Alamos Code RAGE is an adaptive mesh refinement (AMR) radiative hydrodynamical code with a second-order conservative Godunov hydro scheme. It can utilize grey and multigroup flux-limited diffusion to model the flow of radiation. The RAGE simulations in our study have a root grid with 100,000 cells and allow up to 4 levels of refinement for up to 16 times more resolution. Opacities for radiation transport are derived from Los Alamos OPLIB database for the diffuse densities ($\sim 10^{-20}$ g/cm$^3$) typical of astrophysical scenarios. Although RAGE has three-temperature physics capability, in which ions, electrons, and photons can all have distinct temperatures, we use two-temperature physics, in which matter and radiation temperatures, though coupled, are evolved separately to better capture shock breakout. For details on RAGE, see Gittings et al. (2008) or Frey et al. (2013) for its application to supernova problems. 

Since Pop III stars are thought to die in low-density HII regions (Whalen et al. 2004) but may be enveloped by a low-density wind following the expulsion of its their hydrogen layer, we join a simple $r^{-2}$ wind density profile with an initial density of $2 \times 10^{-18}$ g/cm$^3$ to the surface of the star with an intervening bridge that has an $r^{-20}$ density profile. The bridge mitigates numerical instabilities in the radiation solution in RAGE that would otherwise arise if the density at the surface of the star were abruptly dropped to that of the diffuse wind. We take the speed of the wind to be 1000 km/s and its composition to be primordial, 76\% hydrogen and 24\% helium by mass. When the wind falls to a number density of 0.1 particles per cubic cm, it is replaced by a uniform density profile similar to that of the ambient HII region.

RAGE runs require about 20,000 hours of CPU time on Los Alamos supercomputers and evolve the explosions out to three years.

\section{Results and Discussion}

The total luminosity (energy/sec) as a function of time is plotted in Figure 3 for our HNe (upper panel) and some of our PI SNe (lower panel). Shock breakout is evident in all 11 events as the brief luminous pulse that lasts for about 1000 s (or about 20 min). They have about the same duration because the stars have similar radii. The peak luminosity increases with explosion energy. Although shock breakout is the brightest stage of the explosions, it will not be visible today. Most of the photons at this moment are x-rays or hard ultraviolet (UV) that are absorbed by neutral hydrogen in the early universe before they can reach earth. Those that are not absorbed would be redshifted into the extreme UV by the time they reach the Milky Way and would be stopped in its outer layers. Rebrightening due to $^{56}$Ni decay is also visible in most of the light curves at $10^6 - 10^7$ seconds, or at about 3 weeks to 3 months. The degree of rebrightening is proportional to the Ni mass, which generally scales with explosion energy. The least energetic SNe exhibit little or no rebrightening because they do not form much Ni.

\begin{figure}
\includegraphics[width = 280 pt]{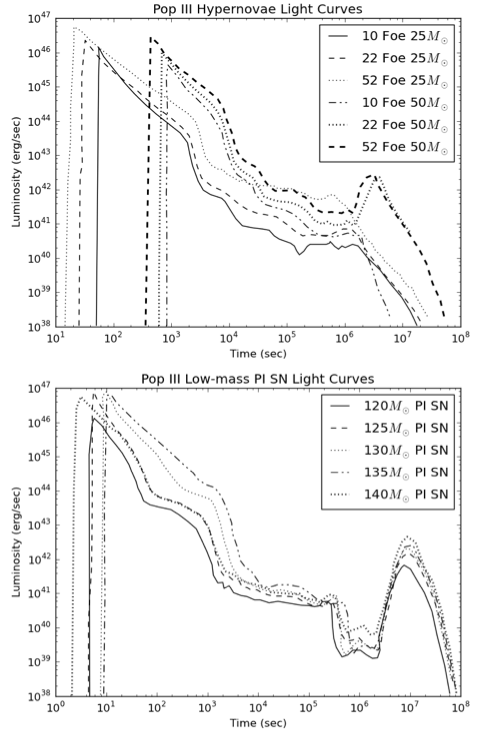}
\caption{\footnotesize The upper panel contains light curves (total luminosity over all wavelengths as a function of time) for our hypernova runs. In the lower panel, we present light curves for the 5 most massive pair instability supernovae (120-140 solar masses). These light curves are calculated in the frame of the star (i.e., they do not account for cosmological redshift and obscuring gas along the line of sight). Note the bump in the light curves around $10^6 - 10^8$ seconds, which is due to radioactive decay of $^{56}$Ni.}
\end{figure}

Future observations will not measure the total luminosities of these events nor are these stars in the local universe, so Figure 3 by itself does not give us much information regarding how observable these events will be today. Observations will instead provide fluxes in specific observing bands in the near infrared (NIR) at 2 - 4 microns. Surveys will hunt for the first SNe in the NIR because any wavelengths in the rest frame of the SN that are shorter than those redshifted into the NIR today will be absorbed by the early universe.

We show NIR light curves for the 50-solar mass 52-foe HN and the 120-solar mass PI SN in Figures 4 and 5. It is clear that HNe will not be visible at redshifts beyond $10 -15$, or about 800 million years after the Big Bang. In Figure 5, we see that PI SNe with masses below 140 solar masses will only be visible at lower redshifts still ($z \sim 3-8$). While such events will not reveal the properties of the first stars, they will probe the stellar populations of the first galaxies, which form at these somewhat lower redshifts. Why are these highly energetic explosions only visible at much lower redshifts than only slightly more energetic 140- to 260-solar mass PI SNe, which can be detected in the first generation of stars? It is primarily because the progenitor has a lower mass and smaller radius at the time of the explosion. The fireball cools at earlier times (and hence smaller radii) and therefore is not as luminous in the bands that are eventually redshifted into the NIR in the Earth frame. We find that this is a general property of highly energetic explosions of compact Pop III stars (Smidt et al. 2014a, Smidt J, Whalen D, Chatzopoulus E, Wiggins B, Fryer C. 2014b, \textit{Astrophysical Journal,} accepted).

\begin{figure}
\includegraphics[width = \columnwidth]{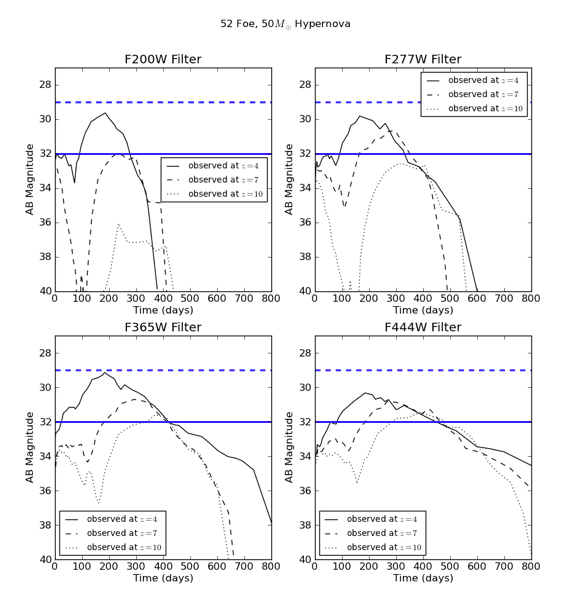}
\caption{\footnotesize Light curves for the 50-solar mass 52-foe hypernova corrected for redshift and absorption by intervening neutral hydrogen from various redshift distances ($z=4$ being the closest and $z=10$ being the most distant). The different panels correspond to 4 NIRcam long-wavelength filters on the James Webb Space Telescope (JWST). The dashed horizontal lines represent the detection limit of Wide-Field Infrared Survey Telescope (WFIRST) after spectrum stacking, and the solid horizontal line is the detection limit for the JWST. This hypernova would, in principle, be visible out to $z=10$ in some filters. }
\end{figure}

\begin{figure}
\includegraphics[width = \columnwidth]{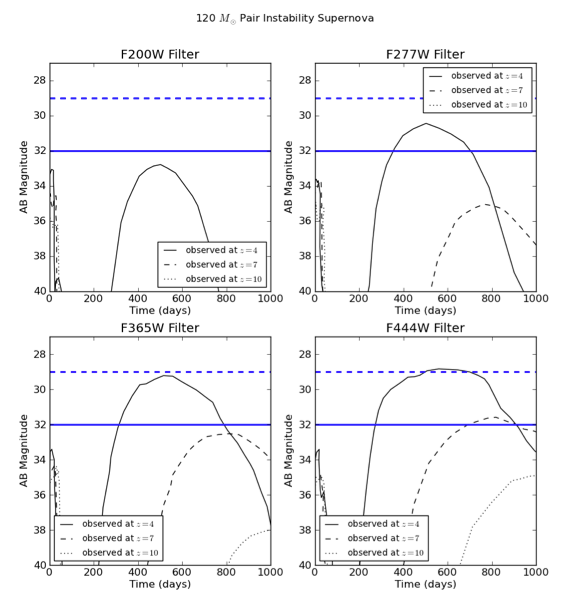}
\caption{\footnotesize Light curves for the 120-solar mass pair instability supernova corrected for redshift and absorption by intervening neutral hydrogen from various redshift distances ($z=4$ being the closest and $z=10$ being the most distant). The different panels correspond to 4 NIRcam long-wavelength filters on the James Webb Space Telescope (JWST). The dashed horizontal line represents the detection limit of Wide-Field Infrared Survey Telescope (WFIRST) after spectrum stacking, and the solid horizontal line is the detection limit for the JWST. }
\end{figure}

But even if a telescope is sensitive enough to detect a primordial SN, there is no guarantee that it will actually come across one in its own lifetime. This depends on the field of view of the instrument and the number of events per square degree on the sky over some interval in redshift. The event rate in turn depends on the Pop III star formation rate. Telescopes like JWST are very sensitive but have very narrow fields of view. NIR missions such as Euclid and the Wide-Field Infrared Survey Telescope (WFIRST) are less sensitive but will survey the entire sky. These instruments could, in principle, harvest large numbers of ancient SNe. 

Perhaps the greatest challenge to detecting Pop III SNe is the low star formation rate during the era of the first stars. Another is that the first SNe are efficient at chemically enriching the early universe. A single PI SNe would spew its heavy elements deep into space, altering the chemistry of large regions of the universe, so there may be a relatively narrow window in redshift in which a SN can be guaranteed to be truly a Pop III event (Wise et al. 2011; Muratov et al. 2013). Both factors limit the total number of Pop III SNe on a given patch of the sky. In lieu of direct observations of Pop III stars, we must rely on cosmological simulations of early star formation for SN rates. Unfortunately, differences in physics between the computer models can cause their predictions of star formation rates to vary by factors of 100 or more (see Whalen et al. 2014a). In particular, Johnson et al (2013b) found that HN rates could be as large as 1000 per year, with most occurring fairly late in the era of the first stars ($z<10$ or about 13.3 billion years ago). Campisi et al. (2011) found a more conservative estimate of about 100 HN events per year across this era. We note that even the failure to detect Pop III SNe in future surveys would be useful because it would rule out the cosmological models with the most optimistic star formation rates.

Could Pop III HNe be found by radio telescopes in surveys? Meiksin and Whalen (2013) have analyzed simulations of Pop III explosions in cosmological halos carried out with the ZEUS-MP code to estimate radio fluxes from HN and core-collapse SN remnants. They find that energetic HNe could be as bright as a few microJanskys in the L and 3-GHz bands, well within the detection limits of existing radio telescopes such as eMerlin and the Jansky Very Large Array (JVLA) Back-of-the-envelope calculations reveal that as many as two radio HNe could be present in a square degree of sky at any given time. To achieve $\sim 3$-microJansky sensitivity in the L band requires the JVLA to dwell on a single region of the sky for nearly 100 hours. Roughly eight such episodes would be required to reject the claim of two hypernovae per square degree with $\sim 95$\% confidence, bringing the total project time for such an undertaking up to a staggering 800 hours on the world?s premier radio telescope. Further, if a supernova candidate were identified, follow-up over the space of years would be required to uniquely identify the event as a primordial explosion. Such surveys are possible, however. We have recently begun collaborations with Chris Hales of the National Radio Astronomy Observatory, who is leading a capabilities test of the VLA with a 1000-hour survey on a single patch of sky in L band that will take place over years. The survey will achieve sensitivities that would otherwise only be attainable by future radio telescope arrays like the Square-Kilometer Array, which will be built in South Africa and be able to detect Pop III core-collapse SNe in addition to HNe. Efforts to find the first cosmic explosions could piggyback on such current surveys. The detection of a primordial supernova will be among the landmark achievements in astronomy in the coming decade.

\section{Conclusion}

In this paper we have described recent efforts by the Los Alamos Light Curve Project to assess the observability of some types of primordial SNe. We have considered the PI SN explosions of compact 90- to 140-solar mass Pop III stars and 25- and 50-solar mass Pop III HNe. We find that these events, although highly energetic, will not be bright enough to be seen at Cosmic Dawn by next-generation telescopes but may be visible in the earliest galaxies. They will complement other types of Pop III SNE as probes of the primeval universe. However, HNe (and core-collapse SNe, but not PI SNe) might be found among the first generation of stars in the radio, and we are currently collaborating with astronomers at the VLA on a 1000-hour capability survey to find them. This survey will be done over several years and will be ideal for detecting and uniquely identifying the remnants of ancient HNe.

\section*{acknowledgements}
BW is grateful for support from the High Impact Doctoral Research Award (HIDRA) at Brigham Young University.  We would like to thank our two anonymous referees for their feedback that greatly enhanced the quality of this review article. We would also like to thank Dr. Bryce Christensen of Southern Utah University for his thoughtful review and suggestions which helped us to prepare this article for broad readership beyond our own field. 


\section*{References}

\footnotesize

\leftskip 0.1in
\parindent -0.1in

Bromm V, Yoshida N, Hernquist L, McKee C. 2009, \textit{Nature}, 49-54

Campisi MA, Maio U, Salvaterra R, Ciardi B. 2011, \textit{Monthly Notices of the Royal Astronomical Society}, 416, 2760

Chatzopoulos E, Wheeler JC. 2012, \textit{Astrophysical Journal,} 748, 42

Christleib N, Bessell M, Beers T, Gustasson B, Korn A, Barklem P, Karlsson T, MizunoWiedner M, Rossi S. 2002, \\textit{Nature,} 419, 904-906

Frebel A, Aoki W, Christlieb N, Ando H, Asplund M, Barklem P, Beers T, Eriksson K, Fechner C, Fujimoto M, Honda S, Kagino T, Minezaki T, Nomoto K, Norris J, Ryan S, Takada-Hidai M, Tsangarides S, Yoshii Y. 2005, \textit{Nature,} 434, 871-873

Frey L, Even W, Whalen D, Fryer C, Hungerford A, Fontes C, Colgan J. 2013, \textit{Astrophysical Journal,} 204, 16

Fryer C, Benz W, Herant M, Colgate SA. 1999, \textit{Astrophysical Journal,} 516, 892-899

Gittings M, Weaver R, Clover M, Betlach T, Byrme N, Coker R, Dendy E, Hueckstaedt R, New K, Oakes W, Ranta D, Stefan R. 2008, arXiv, 0804.1394.

Hirano S, Hosokawa, Yoshida N, Umeda H, Omukai K, Chiaki G, Yorke H. 2014, \textit{Astrophysical Journal,} 781, 60-81

Hosokawa T, Omukai K, Yoshida N, Yorke H. 2011, \textit{Science,} 334, 1250-1253

Joggerst C C, Almgren A, Bell J, Heger A, Whalen D, Woosely S E, 2009, \textit{Astrophysical Journal,} 709, 11-26

Johnson JL, Dalla VC, Khochfar S. 2013, \textit{Monthly Notices of the Royal Astronomical Society,} 428, 1857

Magee NH, Abdallah Jr J, Clar REH, Cohen JS, Collins LA, Csanak G, Fontes G, Gauger A, Keady JJ, Kilcrease DP, Merts AL. 1995, \textit{Astronomical Society of the Pactific Conference Series,} Vol 78, Astrophysical Applications of Powerful New Databases, ed. SJ Adelman, WL Wiese, 51

Meiksin A, Whalen D. 2013, \textit{Monthly Notices of the Royal Astronomical Society,} 430, 2854-2863

Moretti A, Ballo L, Braito V, Caccianiga A, Ceca R, Gilli R, Salvanterra R, Severgnini P, Vignali C. 2014, \textit{Astronomy and Astrophysics,} 563, A46

Muratov A, Gnedin O, Gnedin N, Zemp M. 2013, \textit{Astrophysical Journal,} 773, 19

Nomoto K, Iwamoto K, Mazzali PA, Umeda H, Nakamura T, Patat F, Danziger IJ, Young TR, Suzuki T, Shigeyama T, Augusteijn T, Doublier V, Gonzalez JF, Boehnhardt H, Brewer J, Hainaut OR, Lidman C, Leibundgut B, Cappellaro E, Turatto M, Galama TJ, Vreeswijk PM, Kouveliotou C, Van Paradijs J, Pian E, Palazzi E, Frontera F. 1998, \textit{Nature,} 395, 672674

Pan T, Loeb A, Kasen D. 2012, \textit{Monthly Notices of the Royal Astronomical Society,} 423, 2203

Paxton B, Cantiello M, Arras P, Bildsten L, Brown E, Dotter A, Mankovich C, Montgomery MH, Stello D, Timmes FX, Townsend R. 2013, Astrophysical Journal Supplement, 192, 3
Rakavy G, Shaviv G. 1967, \textit{Astrophysical Journal,} 148, 803

Regan J, Haehnelt M. 2009a, \textit{Monthly Notices of the Royal Astronomical Society,} 393, 858-871

Smidt J, Whalen D, Even W, Wiggins B, Johnson J, Fryer C, Stiavelli M. 2014a, \textit{Astrophysical Journal,} 797, 97

Smidt J, Whalen D, Chatzopoulos E, Wiggins B, Chen K-J, Kozyreva A, Even W, 2014b, arXiv:1411.5377 (accepted to the \textit{Astrophysical Journal})

Weaver TA, Zimmermann GB, Woosley SE. 1978, \textit{Astrophysical Journal,} 225, 1021

Whalen D, Abel T, Normon M. 2004, \textit{Astrophysical Journal,} 610, 14

Whalen D. 2012, arXiv, 1209.4688

Whalen D, Even W, Frey L, Smidt J, Johnson J, Lovekin C, Fryer C, Stiavelli M, Holz D, Heger A, Woosley SE, Hungerford A. 2013a, \textit{Astrophysical Journal,} 777, 110

Whalen D, Joggerts C, Fryer C, Stiavelli M, Heger A, Holtz D. 2013b, \textit{Astrophysical Journal,} 768, 95

Whalen D, Smidt J, Johnson J, Holz D, Stiavelli M, Fryer C. 2014a, arXiv 1312.6330

Whalen D, Smidt J, Even W, Woosely SE, Heger A, Stiavelli M, Fryer C. 2014b, \textit{Astrophysical Journal,} 781, 106

Wise J, Turk M, Abel T. 2008, \textit{Astrophysical Journal,} 682, 745-757

Wise J, Turk M, Norman M, Abel T. 2011, \textit{Astrophysical Journal,} 745, 50

Woosley SE. 1993, \textit{Astrophysical Journal,} 405, 273

Young P A, Fryer C L, 2007, \textit{Astrophysical Journal,} 664, 1033


\end{document}